\documentclass[12pt]{article}
\newcommand{\vecbo}[1]{\mbox{\boldmath $#1$}}
\begin{document}
\title{\bf QED vacuum between an unusual \\pair of plates}

\author{M. V. Cougo-Pinto\thanks{e-mail:{\it marcus@if.ufrj.br}}, C.
Farina\thanks{e-mail:{\it farina@if.ufrj.br}}, F. C.
Santos\thanks{e-mail:{\it filadelf@if.ufrj.br}} \\ and \\ A. C.
Tort\thanks{e-mail:{\it tort@if.ufrj.br}}\\
\\{\small\it Instituto de F\'{\i}sica, 
Universidade Federal do Rio de Janeiro}\\
{\small\it Caixa Postal 68528, 21945-970 - Rio de Janeiro, RJ,  Brazil}}
\date{\today}
\maketitle
\vfill
\begin{abstract}{We consider the photon field between an unusual
configuration of infinite parallel plates: a perfectly conducting plate
$(\epsilon\to\infty)$ and an infinitely permeable one $\mu\to\infty)$. After
quantizing the vector potential in the Coulomb gauge, we obtain explicit
expressions for the vacuum expectation values of field operators of the form  
$<{\hat E}_i{\hat E}_j>_0$ and $<{\hat B}_i{\hat B}_j>_0$. These field
correlators allow us to reobtain the Casimir effect for this set up and to
discuss the light velocity shift caused by the presence of plates
(Scharnhorst effect \cite{Scharnhorst,Barton,BarScharn}) for both scalar and
spinor QED.}

\end{abstract}
\vspace{2cm}
{\it PACS numbers}: 11.10.-z, 11.10.Mn\hfill\break
{\it Key words: Casimir effect, Scharnhorst effect, non-linear
electrodynamics}\hfill\break
\vfill

\eject
\section{\bf Introduction}
Ordinary QED deals with processes in unbounded spacetime, with no boundary
conditions whatsoever or external fields imposed on and without
compactification of any spatial dimension. Nonetheless, a number of physical
interesting processes involving photons and electrons (bound or not) occur
within the confines of physical boundaries, that is, within a cavity. As an
example consider the spontaneous emission by an atom. This process is due to
the coupling of electromagnetic vacuum oscillations to the bound electron in
the atom and in free space is a position-independent observable. However,
inside a cavity the vacuum electromagnetic field modes can change
substantially and as a consequence the spontaneous emission rate is affected
and can become position-dependent \cite{KlepHaroche,Milonni73,Philpott73}
(see also the textbook by Milonni \cite{Milonnibook} and references
therein). For a \lq\lq cavity'' comprised by a single metallic wall, for
instance, the spontaneous emission rate goes with the reciprocal of the
fourth power of the distance of the atom to the wall. In a broader sense, we
can say that inside the cavity we can think of the atom as probing the local
fluctuations of the electromagnetic vacuum.

The influence of the atom-cavity interaction on the atomic spontaneous
emission rate is one among a large number of effects of the so-called cavity
QED, a specific branch of QED that basically deals with the influences of
the surroundings of a physical system on its radiative properties (see
ref(s) \cite{Haroche,Berman} for recent reviews). Although the first cavity
QED effect is attributed to Purcell \cite{Purcell}, who pointed out that the
spontaneous emission process associated with nuclear magnetic moment
transitions at radio frequencies could be enhanced if the system were
coupled to a ressonant external electric circuit, we can say that the first
detailed papers on this subject were those written by Casimir and Polder
\cite{Polder} in which, among other things, forces between polarizable atoms
and metallic walls were treated, and by Casimir in his seminal work
\cite{CASIMIR}. In its electromagnetic version, the Casimir effect is the
macroscopic attraction force between two parallel perfectly conducting
infinite surfaces due to the redistribution of normal modes of the vacuum
electromagnetic field between them. Experimentally, the Casimir effect
between metallic surfaces was first observed by Sparnaay \cite{SPARNAAY} and
recently with remarkable accuracy by Lamoreux \cite{LAMO} and Mohideen and
Roy \cite{MOHIDEEN}.  The various Casimir effects have been the subject of
many studies, for a review see \cite{MosteTrunov, PLUNIEN}.

Still another spectacular instance of cavity QED is the Scharnhorst effect
\cite{Scharnhorst,Barton}. This effect is basically the velocity shift
caused by the change in the zero-point energy density of the quantized
electromagnetic field induced by the presence of Casimirlike plates. Recall
that an external electromagnetic field such as that of a propagating light
couples to the quantized radiation field through fermionic loops. The
Scharnhorst effect is not the only example where non-trivial vacua affects
the speed of light. In fact this subject has attracted the attention of many
physicists in the last years
\cite{Adler,Drummond,DanielsShore,Shore,Latorre,Dittrich}. 

It is clear from what was stated above that an analysis of the QED vacuum
inside cavities is crucial for an understanding of its observable
properties. Here we shall consider the QED vacuum confined by an unsual pair
of mirrors. Specifically, we shall place an infinite perfectly conducting
($\epsilon\to\infty$) surface parallel to a second infinite perfectly
permeable ($\mu\to\infty$) surface held at fixed distance $L$ from the
first. This setup was first considered by Boyer in order to compute the
corresponding Casimir effect in the framework of random electrodynamics
\cite{BOYER} and leads to a repulsive force. This result is somewhat
intriguing, since it seems to contradict the explanation given for the usual
attractive Casimir effect which suggests that there is a greater number of
modes outside the plates than inside \cite{Milonnibook}. In fact, this is
not true: there is only a rearrangement of modes, for a nice explanation of
this problem see \cite{Hushwater}. For the generalized $\zeta$-function
approach applied to the repulsive Casimir effect for parallel plates
geometry see \cite{Andre,SANTOS}.

This paper is organized as follows: in section {\bf 2} we determine the
photon field $\vecbo{A}(\vecbo{r},t)$ in the region between Boyer's plates
making use of the Coulomb gauge. Next we also evaluate the field operator
correlators $<{\hat E}_i{\hat E}_j>_0$ and $<{\hat B}_i{\hat B}_j>_0$ with
the aid of a simple but efficient regularization prescription. In section
{\bf 3} we apply our results to reobtain the repulsive Casimir pressure of
this setup. In section {\bf 4} we discuss the Scharnhorst effect but for
this different situation. In particular, we show that, contrary to the case
with of the usual pair of Casimir plates considered by Scharnhorst
\cite{Scharnhorst} and Barton \cite{Barton}, Boyer's plates lead to a
decrease in the speed of a light for propagation perpendicular to the
plates. In section {\bf 5} we discuss the Scharnhorst effect for the case of
scalar QED trying to keep as much as possible a close analogy with the
spinorial QED case. Section {\bf 6} is left for the final remarks and
conclusions.

We use natural units so that Planck's constant $\hbar$ and the speed of
light $c$ are set equal to one. For the electromagnetic fields we employ the
unrationalized gaussian system. The fine structure constant reads $\alpha =
e^2\approx 1/137$.  
\section{\bf Vacuum electromagnetic field between 
Boyer's \\ plates}

The setup we will consider consists of two infinite parallel surfaces (the
plates) one of which will be considered to be a perfect conductor
($\epsilon\to\infty$) while the other is supposed to be perfectly permeable
($\mu\to\infty$). Also, we will choose Cartesian axes in such a way that the
axis ${\cal OZ}$ is perpendicular to both surafces. The perfectly conducting
surface will be placed at $z=0$ and the permeable one at $z=L$. The
electromagnetic fields must satisfy the following boundary conditions: {\bf
(a)} the tangential components $E_x$ and $E_y$ of the electric field as well
as the normal component $B_z$ of the magnetic field must vanish on the
metallic plate at $z=0$. ({\bf b)} The tangential components $B_x$ and $B_y$
of the magnetic field must vanish on the permeable plate at $z=L$. It is
convenient to work with the vector potential $\vecbo{A}(\vecbo{r},t)$ in the
Coulomb gauge in which $\vecbo{\nabla\cdot A}(\vecbo{r},t)=0$,
$\vecbo{E}(\vecbo{r},t)=-\partial\vecbo{A}(\vecbo{r},t)/\partial t$ and
$\vecbo{B}(\vecbo{r},t)=\vecbo{\nabla\times}\vecbo{{A}(\vecbo{r},t)}$. Then
the physical boundary conditions combined with our choice of gauge permit us
to translate the boundary conditions in terms of the vector potential
components. 
At $z=0$ we have:
\begin{equation}
A_x(x,y,0,t)=0\,;\;\;\;\; A_y(x,y,0, t)=0\,;\;\;\;\;{\partial \over\partial
z}A_z(x,y,0,t)=0\,,
\end{equation}
On the other hand, at $z=L$ we have:
\begin{equation}
{\partial\over\partial x} A_x(x,y,L, t)=0\,;\;\;\;\; {\partial \over\partial
y}A_y(x,y,L,t)=0\,;\;\;\;\; A_z(x,y,L,t)=0\,.
\end{equation}
The appropriate vector potential $\vecbo{A}(\vecbo{r},t)$ that satisfies the
wave equation, the Coulomb gauge condition and the previous boundary
conditions can be written in the form:
\begin{eqnarray}\label{VECPOT}
\mbox{\bf A}(\mbox{\bf r},t) &=& {1\over \pi}\left({\pi\over
L}\right)^{1\over 2}\sum_{n=0}^\infty
\int\,{d^2\kappa\over\sqrt{\omega}} \left\{a^{(1)}(\vecbo{\hat\kappa}, n)
\vecbo{\hat\kappa}\times\hat{\mbox{\bf
z}}\sin\left[\left(n+{1\over2}\right){\pi z\over L}\right]
\right.\nonumber\\
&+& \left. a^{(2)}(\vecbo{\kappa}, n)\left[\vecbo{\hat\kappa}
{i(n+{1\over2})\over \omega  L} \sin\left[\left(n+{1\over2}\right){\pi
z\over L}\right]
-\vecbo{{\hat z}}
{\kappa\over\omega}\cos\left[\left(n+{1\over2}\right){\pi z\over
L}\right]\right]\right\}e^{i(
\vecbo{\kappa}\cdot
\vecbo{\rho}-\omega t)}\;\nonumber\\
&+& \mbox{Hermitian conjugate}\; ,
\end{eqnarray}
where $\vecbo{\kappa} = (k_x, k_y)$ and $\vecbo{\rho}$ is the position
vector in the $xy$-plane. The normal frequencies are given by
\begin{equation}
\omega =\omega(\vecbo{\kappa},n)=\sqrt{\vecbo{\kappa}^2+\left(n+{1\over
2}\right)^2{\pi^2\over L^2}}\,.
\end{equation}
The Fourier coefficients $a^{(\lambda)}(\vecbo{\kappa}, n)$ where
$\lambda=1,2$ is the polarization index, are operators acting on the photon
state space and satisfy the commutation relation
\begin{equation}
\left[a^{(\lambda)}(\vecbo{\kappa},
n),a^{(\lambda^\prime)}(\vecbo{\kappa}^\prime,
n^\prime)\right]=\delta_{\lambda\lambda^\prime}\delta_{nn^\prime}\delta\left
(\vecbo{\kappa}-\vecbo{\kappa}^\prime\right)\,.
\end{equation}
It is convenient to write the vector potential in the general form:
\begin{equation}
\vecbo{A}(\vecbo{r},t)=\sum_\alpha a^\alpha (0)\vecbo{A}^\alpha
(\vecbo{r})e^{-i\omega_\alpha t} +\mbox{H. c.}\,,
\end{equation}
where $\vecbo{A}^\alpha (\vecbo{r})$ denotes the mode functions. The mode
functions for each polarization state obey the Helmholtz equation and
satisfy the boundary conditions stated above. In our case they are given by:
\begin{equation}\label{AUM}
\vecbo{A}_{\vecbo{\kappa}n}^{(1)} (\vecbo{r})={1\over\pi}\left({\pi\over
L}\right)^{1\over 2}{1\over\omega^{1\over
2}}\sin\left[\left(n+{1\over2}\right){\pi z\over
L}\right]e^{-i\vecbo{\kappa\cdot\vecbo{\rho}}}\,\vecbo{\hat\kappa\times\hat
z} \,,
\end{equation}
and
\begin{equation}\label{ADOIS}
\vecbo{A}_{\vecbo{\kappa}n}^{(2)} (\vecbo{r})={1\over\pi}\left({\pi\over
L}\right)^{1\over 2}{1\over\omega^{1\over
2}}\left[\vecbo{\hat\kappa}{in\pi\over
L\omega}\sin\left[\left(n+{1\over2}\right){\pi z\over L}\right]-\vecbo{\hat
z}{\kappa\over\omega}\cos\left[\left(n+{1\over2}\right){\pi z\over
L}\right]\right]e^{-i\vecbo{\kappa\cdot\vecbo{\rho}}}\,.
\end{equation}
Next we evaluate the electric field operator $\vecbo{E}(\vecbo{r},t)$.
Recalling that $a^\alpha |0>=0$ and ${a^\alpha}^\dagger |0>=0$ we first
write for the correlators $<E_i(\vecbo{r},t)E_j(\vecbo{r},t)>_0$ a general
expression of the form:
\begin{equation}\label{E0}
<E_i(\vecbo{r},t)E_j(\vecbo{r},t)>_0=\sum_\alpha
E_{i\alpha}(\vecbo{r})E_{j\alpha}^*(\vecbo{r})\,.
\end{equation}
In our case (\ref{AUM}) and (\ref{ADOIS}) yield
\begin{equation}\label{E1}
\vecbo{E}_{i\vecbo{\kappa}n}^{(1)}
(\vecbo{r})={i\over\pi}\left({\pi\over L}\right)^{1\over
2}{1\over\omega^{1\over 2}}\sin\left[\left(n+{1\over2}\right){\pi z\over
L}\right]e^{-i\vecbo{\kappa\cdot\vecbo{\rho}}}\,
(\vecbo{\hat\kappa\times\hat z})_i \,,
\end{equation}
and
\begin{equation}\label{E2}
\vecbo{E}_{i\vecbo{\kappa}n}^{(2)} (\vecbo{r})={i\over\pi}\left({\pi\over
L}\right)^{1\over 2}{1\over\omega^{1\over
2}}\left[\vecbo{\hat\kappa}_i{in\pi\over
L\omega}\sin\left[\left(n+{1\over2}\right){\pi z\over L}\right]-\vecbo{\hat
z}_i{\kappa\over\omega}\cos\left[\left(n+{1\over2}\right){\pi z\over
L}\right]\right]e^{-i\vecbo{\kappa\cdot\vecbo{\rho}}}\,.
\end{equation}
Now we substitute (\ref{E1}) and (\ref{E2}) into (\ref{E0}), write
$\vecbo{\hat\kappa}_i=\cos\phi\,\delta_{ix}+\sin\phi\,\delta_{iy}$,
$\vecbo{\hat z}_i=\delta_{iz}$ and $(\vecbo{\hat\kappa\times\hat
z})_i=\sin\phi\,\delta_{ix}-\cos\phi\,\delta_{iy}$, where $\phi$ is the
azimuthal angle in the $xy$-plane and compute all angular integrals. In this
way we wind up with
\begin{eqnarray}\label{VEV}
&&<{\hat E}_i(\vecbo{r},t){\hat E}_j(\vecbo{r},t)>_0 \;\;=\;\; \left({2\over
\pi}\right)\left({\pi\over L}\right){\delta^{\|}_{ij}\over
2}\sum_{n=0}^\infty\sin^2{\left[\left(n+{1\over 2}\right){\pi z\over
L}\right]}\int_0^\infty d\kappa\,\kappa\,\omega (\vecbo{\kappa}, n) \nonumber \\
&+& \left({2\over \pi}\right)\left({\pi\over L}\right)\left({\pi\over
L}\right)^2{\delta^{\|}_{ij}\over
2}\sum_{n=0}^\infty\sin^2{\left[\left(n+{1\over2}\right){\pi z\over
L}\right]}\left( n+{1\over2}\right)^2 \int_0^\infty
d\kappa\,\kappa\,\omega^{-1} (\vecbo{\kappa}, n) 
\nonumber \\
&+& \left({2\over \pi}\right)\left({\pi\over
L}\right)\delta^{\bot}_{ij}\sum_{n=0}^\infty\cos^2{\left[\left(n+{1\over2}
\right){\pi z\over L}\right]}\int_0^\infty d\kappa\,\kappa^3\,
\omega^{-1}(\vecbo{\kappa}, n)\,,
\end{eqnarray}
where $\delta^{\|}_{ij}:=\delta_{ix}\delta_{jx}+\delta_{iy}\delta_{jy} $
and $\delta^{\bot}_{ij}:=\delta_{iz}\delta_{jz}$. The previous equation is
only a formal expression for the field correlator 
$\langle {\hat E}_i(\vecbo{r},t){\hat E}_j(\vecbo{r},t)\rangle _0$, since it
is an ill-defined expression plagued by divergent terms. Therefore, it lacks
of physical meaning unless we adopt a regularization prescription. We will
first regularize the integrals in equation (\ref{VEV}) by using a method
based on analytical extension in the complex plane. The idea is the
following: take for example the first integral  that appears on the r.h.s.
of (\ref{VEV}),
$${\cal I}_1(n,L):=\int_0^\infty d\kappa \kappa\left(
\kappa^2+{(n+{1\over2})^2\pi^2\over L^2}\right)^{1/2}\; . $$

\noindent
Since this integral diverges for large $\kappa$, it is natural to modify the
integrand so that the integral becomes finite. Our choice will be simply
$$ {\cal I}_1(n,L)\longrightarrow {\cal I}^{reg}_1(n,L;s):= 
\int_0^\infty d\kappa \kappa\left( \kappa^2+{(n+{1\over2})^2\pi^2\over
L^2}\right)^{1/2-s} \; $$
and after the calculations we will take the limit $s\to 0$. For the moment,
let us assume that $\Re\, s$ is large enough to give a precise mathematical
meaning for the previous integral. Then, making use of the following
integral representation of the Euler beta function, {\it c.f.} formula {\bf
3.251}.2 \cite{Grad}:
\begin{equation}
\int_0^\infty dx\,x^{\mu-1}\left(x^2+a^2\right)^{\nu-1} =
{B\over2}\left({\mu\over2},1-\nu-{\mu\over2}\right) a^{\mu+2\nu-2}\,,
\end{equation}
where $B(x,y)=\Gamma(x)\Gamma(y)/\Gamma(x+y)$, which holds for
$\Re\,\left(\nu+{\mu\over 2}\right)< 1$ and $\Re\,\mu>0$, we get
\begin{equation}
{\cal I}^{reg}_1(n,L;s)={1\over2}\left[ (n+{1\over2}{\pi\over L}
\right]^{3-2s}\; {\Gamma(s-3/2)\over \Gamma(s-1/2)}
={1\over (2s-3)}\left[(n+{1\over2}{\pi\over L}
\right]^{3-2s} 
\end{equation}

\noindent
Inserting this result into the first term of the r.h.s. of (\ref{VEV}) (call
it ${\cal T}_1)$, it takes the form:
\begin{equation}
{\cal T}_1=\left({1\over 2s-3}\right)\left({\pi\over L}\right)^{3-2s}
{\delta^{\|}_{ij}\over 2L}\left\{ \zeta_H(2s-3,1/2)-\sum_{n=0}^\infty 
\left( n+{1\over 2}\right)^{3-2s}\cos\left[ {2(n+1/2)\pi z\over L}\right]
\right\}\; ,
\end{equation}
where $\zeta_H(z,a)$ is the well known Hurwitz zeta function. Making the
analytical extension to the $s$-complex plane and taking the limit
$s\rightarrow 0$, we get
\begin{equation}
{\cal T}_1=-{1\over 6\pi}\left({\pi\over L}\right)^4
\delta^{\|}_{ij}\left\{\left(-{7\over 8}\right)\times 
{1\over 120}-G(\pi z/L)\right\}\; ,
\end{equation}
where we made use of $\zeta_H(-3,1/2)=(-7/8)\times(1/120)$ and defined 
\begin{eqnarray}\label{GFINAL}
G\left(\xi\right)& = & -\frac{1}{8}\times {d^3\over d\xi^3}\left({1\over
2\sin{\xi}}\right) \nonumber \\
& = & \frac{1}{8}\left(3{\cos^3{\xi}\over
\sin^4{\xi}}+\frac{5}{2}{\cos{\xi}\over \sin^2{\xi}}\right)\,.
\end{eqnarray}
\noindent
Analogous calculations can be performed with the other terms of the r.h.s.
of (\ref{VEV}). It is then straightforward to show that
\begin{equation}\label{COREIEJ}
\langle{\hat E}_i(\vecbo{r},t){\hat
E}_j(\vecbo{r},t)\rangle_0=\left({\pi\over L}\right)^4{2\over
3\pi}\left[\left(-{7\over
8}\right)\left(-\delta^{\|}+\delta^{\bot}\right)_{ij}\;{1\over
120}+\delta_{i,j}G(\pi z/L)\right]\,,
\end{equation}
and proceeding in the same way we did in the evaluation of the electric
field correlators  we obtain
\begin{equation}\label{CORBIBJ}
\langle{\hat B}_i(\vecbo{r},t){\hat
B}_j(\vecbo{r},t)\rangle_0=\left({\pi\over L}\right)^4{2\over
3\pi}\left[\left(-{7\over
8}\right)\left(-\delta^{\|}+\delta^{\bot}\right)_{ij}\;{1\over
120}-\delta_{i,j}G(\pi z/L)\right]\,,
\end{equation}

\noindent for the magnetic field correlators. A straightforward calculation
along the lines given here or the use of time-reversal invariance shows that
the correlators $<E_i(\vecbo{r},t)B_j(\vecbo{r},t)>_0=0$. In passing,
observe that no substractions whatsoever were required in our regularization
procedure. This is a common feature of regularization prescriptions based on
the analytical extension. However, other methods where the subtraction of
the field correlators involving no boundary conditions are present can be
used yielding the same results. 
\section{\bf The Casimir effect between Boyer's plates}

In order to get confidence in the previous results for the field operator
correlators between Boyer's plates, let us reobtain Boyer's result
\cite{BOYER} concerning the Casimir effect for this unusual set up. First,
recall that the zero-point energy density $\rho_o$ for the electromagnetic
fields is defined by the following vacuum expectation value:
\begin{equation}\label{Casi}
\rho_0=\frac{1}{8\pi}<\vecbo{E}^2+\vecbo{B}^2>_0\,.
\end{equation}
Making use of (\ref{COREIEJ}) and (\ref{CORBIBJ}) we obtain the
position-dependent correlators:
\begin{equation}
<\vecbo{E}^2>_0=\left({\pi\over L}\right)^4{2\over 3\pi}\left[{7\over
8\times 120}+3G(\xi)\right]\,,
\end{equation}
\begin{equation}
<\vecbo{B}^2>_0=\left({\pi\over L}\right)^4{2\over 3\pi}\left[{7\over
8\times 120}-3G(\xi)\right]\,.
\end{equation}
If we add these two equations the position-dependent terms will cancel out
and if we substitute the result into (\ref{Casi}) we will obtain:
\begin{equation}
\rho_0= {7\over 8}\times {\pi^2\over 720 L^4}\,,
\end{equation}
which is the position-independent and positive Casimir energy density
leading to a repulsive force per unit area between the plates \cite{BOYER,
Andre, SANTOS}.

It is also convenient to analyze the behavior of the correlators
$<\vecbo{E}^2>_0$ and $<\vecbo{B}^2>_0$ in the situations where one of the
plates is removed. Let us first consider the limit of a single metal plate
located at $z=0$. This means that we are taking the limit $L\to\infty$ in
our previous results. The results are: 
\begin{equation}\label{SMLEO}
<\vecbo{E}^2>_0\approx +{3\over 4\pi z^4}\,,
\end{equation}
and
\begin{equation}\label{SMLBO}
<\vecbo{B}^2>_0\approx - {3\over 4\pi z^4}\,,
\end{equation}
in agreement with the literature \cite{LR}
On the other hand, the limit of a single infinitely permeable plate is
obtained by removing the metal plate. This can be accomplished if we
consider the limits $L\to\infty$, $z\to\infty$ in the previous results, but
with $L-z<<L$. For this case we obtain: 
\begin{equation}\label{SMLEL}
<\vecbo{E}^2>_0\approx -{3\over 4\pi (z-L)^4}\,,
\end{equation}
and
\begin{equation}\label{SMLBL}
<\vecbo{B}^2>_0\approx  +{3\over 4\pi (z-L)^4}\,.
\end{equation}
Equations (\ref{SMLEL}) and (\ref{SMLBL}) are new results. Let us turn our
attention now to one of the most intriguing properties of the QED vacuum:
its anisotropy and the concomitant consequences on the speed of light.
\section{\bf The Scharnhorst effect for the spinor QED}

The Scharnhorst effect \cite{Scharnhorst,Barton} is basically the light
velocity shift in the QED vacuum caused by the presence of two parallel
plates for propagation inside the plates and perpendicular to them. This was
shown to occur for small frequencies $\omega<<m$ (soft photon approximation)
and in the weak field limit. For the case of metallic plates, Scharnhorst
\cite{Scharnhorst} and later on Barton \cite{Barton} showed that the phase
velocity, which for this case coincides with the phase velocity for small
frequencies, is greater thatn its value in free space ($c$) for propagation
perpendicular to the plates. However, this does not mean that the signal
velocity can be greater than $c$ because to determine the wave front
velocity it is necessary to investigate the dispersion relation in the
infinite frequency limit (see reference 
\cite{Milonni2,BarScharn,MilonniSvozil,Menahem} for some discussion on this
issue). The Scharnhorst effect with a boundary condition other than the
standard one for perfect metallic plates has also been considered
\cite{Bartinho}. It can be understood as follows: the external field as that
describing the propagation of a plane wave interacts with the quantized
electromagnetic fields through the fermionic loops and hence, any change in
the quantized field modes, as for example caused by imposition of boundary
conditions, can in principle modify the wave propagation. In references
\cite{Scharnhorst,Barton} this change was induced by the presence of two
perfect parallel conducting plates. Since these authors assumed that the
plates do not impose any boundary condition on the fermionic field, the
Scharnhorst effect appears only at the two-loop level. Also, because it is a
perturbative effec, it can be obtained by direct computation of the relevant
Feynman diagrams that contribute to the effective action, namely: the two
possible diagrams for the photon polarization tensor at two-loop level. This
was precisely Scharnhorst's approach, who after using a previous
representation for the photon propagator between two metallic plates
obtained by Bordag, Robaschik and Wieczorek \cite{BORDAG} found for
propagation perpendicular to the plates that
\begin{equation}
v_\bot=1+{11\pi^2\over 2^2.3^4.5^2}{\alpha^2\over (mL)^4}\; .
\end{equation}

Later on, the same result was rederived by Barton \cite{Barton} in a more
economic way, where the connection to the Casimir energy density is more
apparent. The starting point in Barton's approach is the addition to the
electromagnetic field lagrangian density of a correction term represented by
the Euler-Heisenberg \cite{HL} effective lagrangian density, so that the
full lagrangian density reads:
\begin{eqnarray}\label{FULLL}
{\cal L}& = & {\cal L}^{(0)}+{\cal L}^{(1)}\nonumber \\
& = & {1\over 8\pi}\left(\vecbo{E}^2-\vecbo{B}^2\right)+
g\left[\left(\vecbo{E}^2-\vecbo{B}^2\right)^2+7\left(\vecbo{E\cdot B}
\right)^2\right]\,,
\end{eqnarray}
where $g:=\alpha^2/5\cdot 3^2\cdot 2^3\cdot\pi^2 m^4$. The lagrangian
density represented by (\ref{FULLL}) describes the first vacuum polarization
effects on slowly varying fields for which the condition $\omega \ll m$
holds and is valid only in the weak field approximation. In other words, the
first non-linear effects to Maxwell equations coming from QED are described
by the quartic terms added to the usual Maxwell lagrangian density in the
above formula. The corresponding vacuum polarization $\vecbo{P}$ and
magnetization $\vecbo{M}$ are given by:
\begin{equation}\label{P1}
\vecbo{P}={\partial{\cal L}^{(1)}\over\partial\vecbo{E}}=4g\left
(\vecbo{E}^2-\vecbo{B}^2\right)\vecbo{E}+14g
\left(\vecbo{E\cdot B}\right)
\vecbo{B}\,,
\end{equation}
and
\begin{equation}\label{M1}
\vecbo{M}={\partial{\cal L}^{(1)}
\over\partial\vecbo{B}}=-4g\left(\vecbo{E}^2-\vecbo{B}^2\right)\vecbo{B}
+14g\left(\vecbo{E\cdot B}
\right)\vecbo{E}\,.
\end{equation}

In order to include a radiative correction into the formalism, we can follow
reference \cite{Barton} and rewrite the fields in equations (\ref{P1}) and
(\ref{M1}) as the sum of two parts, one describing the quantized fields and
the other one describing the classical fields, that is, we write:
$\vecbo{E}\to\vecbo{E}_q+\vecbo{E}_c$ and
$\vecbo{B}\to\vecbo{B}_q+\vecbo{B}_c$ and substitute into (\ref{P1}) and
(\ref{M1}). This procedure is tantamount to the coupling of the external
fields to the quantized ones by means of the intermediary action of a
fermionic loop.  Keeping only the terms which are linear in the classical
fields, we obtain the following expressions for the electric susceptibilty
$\chi^{(e)}_{ij}$  and magnetic susceptibility $\chi^{(m)}_{ij}$ tensors of
the vacuum:
\begin{equation}\label{DIE}
\chi^{(e)}_{ij}= 4g\left[
<\vecbo{E_q}^2-\vecbo{B_q}^2>_0\delta_{ij}+2<E_{qi}E_{qj}>_0
\right]+14g<B_{qi}B_{qj}>_0\,,
\end{equation}
\begin{equation}\label{PER}
\chi^{(m)}_{ij}=
4g\left[-<\vecbo{E_q}^2-\vecbo{B_q}^2>_0\delta_{ij}+2<B_{qi}B_{qj}>_0
\right]+14g<E_{qi}E_{qj}>_0\,.
\end{equation}
The dieletric and permittivity tensors of the vacuum are:
\begin{equation}
\epsilon_{ij}=\delta_{ij}+4\pi\chi^{(e)}_{ij}=\delta_{ij}+
\Delta\epsilon_{ij}\,,
\end{equation}
\begin{equation}
\mu_{ij}=\delta_{ij}+4\pi\chi^{(m)}_{ij}=\delta_{ij}+\Delta\mu_{ij}\,,
\end{equation}
The vacuum expectation values in (\ref{DIE}) and (\ref{PER}) can be easily
calculated with the correlators given by (\ref{COREIEJ}) and
(\ref{CORBIBJ}). If we do this, we obtain for $\Delta\epsilon_{ij}$ and
$\Delta\mu_{ij}$ the results:
\begin{equation}\label{DIE2}
\Delta\epsilon_{ij}=g\left({\pi\over L}\right)^4{16\over
3}\left[\left(-{7\over
8}\right)\left(-\delta^{\|}+\delta^{\bot}\right)_{ij}\;\left({11\over
120}\right)+3\delta_{ij}G(\xi)\right]\,,
\end{equation}
and
\begin{equation}\label{PER2}
\Delta\mu_{ij}=g\left({\pi\over L}\right)^4{16\over 3}\left[\left(-{7\over
8}\right)\left(-\delta^{\|}+\delta^{\bot}\right)_{ij}\;\left({11\over
120}\right)-3\delta_{ij}G(\xi)\right]\,.
\end{equation}
We can also derive  single plate limits for $\Delta\epsilon_{ij}$ and
$\Delta\mu_{ij}$. Making use of the approximations to $G(\xi)$ in the limits
$\xi\to 0$ and $\xi\to\pi$ we have near the conducting plate at $z=0$:
\begin{equation}
\Delta\epsilon_{ij}=-\Delta\mu_{ij}=18g{\delta_{ij}\over z^4}\,,
\end{equation}
and near the permeable plate at $z=L$:
\begin{equation}
\Delta\epsilon_{ij}=-\Delta\mu_{ij}=-18g{\delta_{ij}\over \left(z-L\right)^4}\,.
\end{equation}

Now, we are interested in the refraction index $n=\sqrt{\epsilon\mu}$ and
its first order shift: 
\begin{equation}\label{INDEX}
\Delta n= \frac{1}{2}\left(\Delta\epsilon+\Delta\mu\right)\,,
\end{equation}
for directions of propagation defined by the cartesian axis. Let us consider
first a plane wave propagating in the $OX$-direction with the electric field
vibrating in the $OZ$-direction. Then $\Delta\epsilon\to\Delta\epsilon_{33}$
and $\Delta\mu\to\Delta\mu_{22}$, and from (\ref{DIE2}),  (\ref{PER2}) and
(\ref{INDEX}) we can easily verify that $\Delta
n=\frac{1}{2}\left(\Delta\epsilon_{33}+\Delta\mu_{22}\right)=0$. We obtain
the same result in all instances in which the propagtion is parallel to the
plane of the plates. As a consequence the speed of light remains unchanged
for propagtion parallel to the plates. Now consider a plane wave propagating
along the $OZ$-axis, perpendicularly to the pair of plates.  Consider the
wave polarized in the $OX$-direction, for instance. Then
$\Delta\epsilon\to\Delta\epsilon_{11}$ and $\Delta\mu\to\Delta\mu_{22}$, and
from (\ref{DIE2}),  (\ref{PER2}) and (\ref{INDEX}) we now obtain:
\begin{eqnarray}
\Delta n_\bot & \approx &
\frac{1}{2}\left(\Delta\epsilon_{11}+\Delta\mu_{22}\right)\nonumber \\
& = & +\frac{7}{8}\times{\alpha^2\over\left(mL\right)^4}{11\pi^2\over
2^2\cdot3^4\cdot 5^2}
\end{eqnarray}
which is the result obtained by Scharnhorst \cite{Scharnhorst} and
reobtained by Barton \cite{Barton} multiplied by the factor $-7/8$. The
speed of light in that direction will be:
\begin{equation}
v_\bot\approx
1-\frac{7}{8}\times{\alpha^2\over\left(mL\right)^4}{11\pi^2\over
2^2\cdot3^4\cdot 5^2}<1\; , 
\end{equation}
as anticipated in the begining of this work. The direction-averaged light
velocity between Boyer's plates also satisfies the unifying formula written
down by Latorre, Pascual and Tarrach \cite{Latorre} for spinor QED which reads 
\begin{eqnarray}
\langle v\rangle=1-{44\alpha^2\over 135 m_e^4}\; \rho_0\; .
\end{eqnarray}
It can be shown that this formula can be obtained in the weak field limit of
Dittrich and Gies' approach to the study of non-trivial vacua
\cite{Dittrich}. We will return to this in the next section.
\subsection{\bf The Scharnhorst effect in scalar QED}

Although the interaction of charged fermions of spin $1/2$ with themselves
and with the photon field is described in a very satisfactory way by spinor
QED, we are not prohibited of thinking on other theories. It may be very
instructive to study other theories that, though not realistic, respect all
important physical principles as for instance, the gauge principle and
relativistic invariance. This is the case of the so-called scalar QED, which
describes charged bosons interacting with themselves and with the radiation
field. Naively, we could think that the interaction between the
pseudoscalars charged mesons $\pi^\pm$ and $K^\pm$ could be described by
scalar QED, but this is not true, mainly because these mesons have an inner
structure and their interaction is dominated by the strong interaction. In
fact, since there are no fundamental charged bosons in Nature, scalar QED is
of limited application. However, scalar QED can be viewed as a toy model in
many situations and hence it may shed some light on interesting physical
processes, as we shall see. Without further apologies, we shall consider in
this section the Scharnhorst effect in the framework of scalar QED.  In the
case of scalar QED the analogue of the Euler-Heisenberg effective lagrangian
reads \cite{Schwinger51}:
\begin{equation}
{\cal L}_0^{(1)}=g_0\left[{7\over
4}\left(\vecbo{E}^2-\vecbo{B}^2\right)^2+\left(\vecbo{E\cdot B}
\right)^2\right]\,,
\end{equation}
with $g_0:=\alpha^2/5\cdot 3^2\cdot 2^5\cdot\pi^2\cdot m_o^4$, where $m_o$
is the mass of the hypothetical charged boson associated with $1$-loop
scalar QED. As before, the polarization $\vecbo{P}$ and the magnetization
$\vecbo{M}$ are defined by equations (\ref{P1}) and (\ref{M1}), and as
before we make use of the substitutions
$\vecbo{E}\to\vecbo{E}_q+\vecbo{E}_c$ and
$\vecbo{B}\to\vecbo{B}_q+\vecbo{B}_c$ and keep only terms linear in the
classical fields to obtain the corrections $\Delta\epsilon_{ij}$ and
$\Delta\mu_{ij}$ to the dielectric and permittivity tensors of the scalar
QED vacuum. The results are
\begin{equation}\label{DIECAS}
\Delta\epsilon_{ij}= 28\pi g_0<\vecbo{E}^2-\vecbo{B}^2>_0\delta_{ij}+56\pi
g_0<E_iE_j>_0 + 8\pi g_0<B_iB_j>_0\,,
\end{equation}
\begin{equation}\label{PERCAS}
\Delta\mu_{ij}= -28g_0<\vecbo{E}^2-\vecbo{B}^2>_0\delta_{ij}+ 56\pi g_0
<B_iB_j>_0 + 8\pi g_0<E_iE_j>_0\,.
\end{equation}
Now we can make use of these results and analyze the speed of light in
confined scalar QED vacuum. Since the Scahrnhorst effect for scalar QED has
never been discussed before, we will evaluate the light velocity shifts for
two cases, to wit, for Casimir's plates and for Boyer's plates. 

\noindent 
{\it Casimir's plates}.  We shall consider the two perfectly conducting
plates at $z=0$ and $z=L$ respectively. Expressions for the electric and
magnetic field correlators for the Casimir's plates can be found in, for
instance, \cite{Barton}, here we merely state the results
\begin{equation}\label{COECAS}
<E_i(\vecbo{r},t)E_j(\vecbo{r},t)>_0=\left({\pi\over L}\right)^4{2\over
3\pi}\left[\left(-\delta^{\|}+\delta^{\bot}\right)_{ij}\;{1\over
120}+\delta_{ij}F(\pi z/L)\right]\,,
\end{equation}
and
\begin{equation}\label{COBCAS}
<B_i(\vecbo{r},t)B_j(\vecbo{r},t)>_0=\left({\pi\over L}\right)^4{2\over
3\pi}\left[\left(-\delta^{\|}+\delta^{\bot}\right)_{ij}\;{1\over
120}-\delta_{ij}F(\pi z/L)\right]\,,
\end{equation}
where $F(\xi)$ is defined by:
\begin{equation}
F(\xi):= -\frac{1}{8}\times {d^3\over d\xi^3}\left({1\over 2}\cot{\xi}\right)\,.
\end{equation}
Now we take (\ref{COECAS}) and (\ref{COBCAS}) into (\ref{DIECAS}) and
(\ref{PERCAS}) and after some simple manipulations we end up with
\begin{equation}
\Delta\epsilon_{ij}={16\over 3}g_0\left({\pi\over
L}\right)^4\left[\left(-\delta^{\|}+\delta^{\bot}\right)_{ij}\;\left({1\over
15}\right)+27\delta_{ij}F(\xi)\right]\,,
\end{equation}
and,
\begin{equation}
\Delta\mu_{ij}={16\over 3}g_0\left({\pi\over
L}\right)^4\left[\left(-\delta^{\|}+\delta^{\bot}\right)_{ij}\;\left({1\over
15}\right)-27\delta_{ij}F(\xi)\right]\,.
\end{equation}

With these results we can now calculate the first correction to the
refraction index $\Delta n$ and, consequently, the correction to the speed
of light between Casimir's plates in scalar QED. As in the corresponding
case of spinor QED, we find that the speed of light parallel to the plates
remains unchanged, but the speed of light perpendicular to the plates is
modified by an amount given by
\begin{equation}\label{CAS}
\Delta v_\bot=-\Delta n=+{16\over 45} g_0\left({\pi\over L}\right)^4>0\,.
\end{equation}

It is interesting to compare this result with the analogous effect that
taked place in spinor QED. Assuming the same charge for the particles
(bosons and fermions), we see  that the ratio between the light velocity
shifts for scalar and usual QED is given by
\begin{equation}
{\Delta v_\bot^b\over \Delta v_\bot}=8\times 
\left({m\over m_o}\right)^4\; .
\end{equation}

{\it Boyer's plates}. Now we repeat the procedure for the unusual pair of
plates that we are discussing here. The electric and magnetic field
correlators we need are given by equations (\ref{COREIEJ}) and
(\ref{CORBIBJ}). Substituting into (\ref{DIECAS}) and (\ref{PERCAS}) we obtain
\begin{equation}
\Delta\epsilon_{ij}={16\over 3}g_0\left({\pi\over
L}\right)^4\left[\left(-{7\over
8}\right)\left(-\delta^{\|}+\delta^{\bot}\right)_{ij}\;\left({1\over
15}\right)+27\delta_{ij}G(\xi)\right]\,,
\end{equation}
and,
\begin{equation}
\Delta\mu_{ij}={16\over 3}g_0\left({\pi\over L}\right)^4\left[\left(-{7\over
8}\right)\left(-\delta^{\|}+\delta^{\bot}\right)_{ij}\;\left({1\over
45}\right)-27\delta_{ij}G(\xi)\right]\,.
\end{equation}
Hence, the speed of light between Boyer's plates in the direction
perpendicular to the plates is modified by the amount 
\begin{equation}\label{BO}
\Delta v_\bot=-\Delta n=-{7\over 8}\times {16\over 45} g_0\left({\pi\over
L^4}\right)<0\,.
\end{equation}
The results given by equations (\ref{CAS}) and (\ref{BO}) can be unified by
considering the average taken over all directions of propagation of the
speed of light between the plates. To acomplish this first we write, for
instance, for Casimir's plates: 
\begin{equation}
v(\theta)=1-{16\over 45}g_0\left({\pi\over L}\right)^4
\cos^2{\theta}\,,
\end{equation}
where $\theta$ is the angle between the direction of propagation and the
${\cal OZ}$-axis. Next we take the average over all direction. The result is:
\begin{eqnarray}\label{average}
\langle v\rangle={1\over 4\pi}\oint v(\theta)d\Omega&=&1+{8\alpha^2\over 135
m_o^4}\left({\pi^2\over 720 L^4}\right)\nonumber\\
&=&1+{8\alpha^2\over 135 m_0^4}\; \rho_0\; .
\end{eqnarray}

\noindent
The same result can be obtained from equation (\ref{BO}) with
$\rho_0=(-7/8)\times (\pi^2/720 L^4)$.
This is the scalar QED version of the unifying formula obtained by Pascual,
Latorre and Tarrach for spinor QED \cite{Latorre}, and also as in the spinor
QED case, it corresponds to the  weak field limit of a more general approach
due to Dittrich and Gies \cite{Dittrich}. 
\subsection*{\bf Final remarks}
In this work we have endeavoured to give another example of the consequences
of imposing boundary conditions on QED vacuum oscillations by discussing
these oscillations confined by a somewhat unusual pair of plates. In
particular, we have obtained through a simple regularization procedure the
correlators for the vacuum oscillations of the electromagnetic field
sandwiched between these plates, the associated Casimir energy density and
the natural converse of the original Scharnhorst effect at zero temperature.
Incidentally, observe that contrary to the case of Casimir's plates, in the
case we discussed here there is no critical temperature for which the
Scahrnhorst effect would vanish. Also, as in the case of Casimir's plates,
the refraction index is frequency-independent, for the Euler-Heisenberg
lagrangian density holds only for slowly varying fields. We have also
examined the scalar QED version of the Scharnhorst effect and produced a a
formula that plays the role of the unifying formula due to Latorre, Pascual
and Tarrach for the case of spinor QED.
\subsection*{\bf Acknowledgments}
Two of us (M. V. Cougo-Pinto and C. Farina) wish to acknowledge the partial
financial support of the Conselho Nacional de Pesquisas (CNPq), the
Brazilian research agency.
\clearpage


\begin{thebibliography}{99}
\bibitem{Scharnhorst} Scharnhorst K, (1990) {\it Phys. Lett.} {\bf B236}, 354 
\bibitem{Barton} Barton G, (1990) {\it Phys. Lett.} {\bf B237}, 559
\bibitem{BarScharn} Barton G and Scharnhorst K, (1993) {\it J. Phys.} {\bf
A26}, 2037
\bibitem{KlepHaroche} Haroche S and Klepner D,  (1989) {\it Physics Today}
January, 25 
\bibitem{Milonni73} Milonni P W and Knight P L, (1973) {\it Opt. Commun.}
{\bf 9}, 119.
\bibitem{Philpott73} Philpott M R, (1973) {\it Chem. Phys. Lett.} {\bf 19}, 435
\bibitem{Milonnibook} Milonni P W, (1994) {\it The Quantum Vacuum}, Academic
Press, Boston 1994
\bibitem{Hushwater} Hushwater V, (1997) {\it Am. J. Phys.} {\bf 65}, 381
\bibitem{Haroche} Haroche S, in {\it Fundamental Systems in Quantum Optics},
Les Houches Summer School, Session LIII, edited by J. Dalibard , J.-M.
Raymond, and J. Zinn-Justin (North-Holland, Amsterdam,1992).
\bibitem{Berman} Berman P R, ed (1994) {\it Cavity Quantum Electrodynamics},
Academic Press, Boston
\bibitem{Purcell} Purcell E M,(1946) {\it  Phys. Rev.} {\bf 69}, 681
\bibitem{Polder} Casimir H B G and Polder D, (1948) {\it Phys. Rev.} {\bf
73}, 360
\bibitem{CASIMIR} Casimir H B G, (1948) {\it Proc. Nederl.Wetensch}. {\bf
51}, 793.
\bibitem{SPARNAAY} Sparnaay M J, (1958) {\it Physica} {\bf 24}, 751.
\bibitem{LAMO} Lamoreaux S K, (1997) {\it Phys. Rev. Lett.} {\bf 79}, 5.
\bibitem{MOHIDEEN} Mohideen U and Roy A, {\it A precision measurement of the
Casimir force from $0.1$ to $0.9$ $\mu$m}, hep-ph/9805038
\bibitem{MosteTrunov} Mostepanenko V M and Trunov N N, (1988) {\it Sov.
Phys. Usp.} {\bf 31}, 965
\bibitem{PLUNIEN} Plunien G,  M\"uller B and Greiner W, (1987) {\it Phys.
Rep.} {\bf 134}, 664 
\bibitem{Adler} Adler S L, (1971) {\it Ann. Phys.} NY {\bf 67}, 599
\bibitem{Drummond} Drummond I T and Hathrell S J, (1980) {\it Phys. Rev.}
{\bf D22}, 343
\bibitem{DanielsShore} Daniels R D and Shore G M, (1994) {\it Nuc. Phys.}
{\bf B425}, 634
\bibitem{Shore} Shore G M, (1996) {\it Nuc. Phys.} {\bf B460}, 379
\bibitem{Latorre} Latorre J I, Pascual P and Tarrach R, (1995) {\it Nuc.
Phys.} {\bf B437}, 60
\bibitem{Dittrich} Dittrich W and Gies H (1999) {\it Phys. Lett.} {\bf D58}, ???
\bibitem{BOYER} Boyer T H, (1974) {\it Phys. Rev.} {\bf A9}, 2078
\bibitem{LR} L\"utken C A and Ravndal F, (1985) {\it Phys. Rev. A} {\bf 31},
2082
\bibitem{Andre} M. V. Cougo-Pinto, C. Farina and A. Ten\'orio, {\it Zeta
Function Method for the Repulsive Casimir Effect}. To appear in {\it Braz.
J. Phys.}
\bibitem{SANTOS} F. C. Santos, A. Ten\'orio and A. C. Tort,  {\it Zeta
Function Method for Repulsive Casimir Forces at Finite Temperature},
hep-th/9807162
\bibitem{MilonniSvozil} Milonni P W, Svozil K, (1990) {\it Phys. Lett. }
{\bf B248}, 437
\bibitem{Milonni2} Milonni P W, Mei-Lei Shih, (1992) {\it Contemp. Phys.}
{\bf 33}, 313
\bibitem{Menahem} Ben-Menahem S, (1990) {\it Phys. Lett.} {\bf B250}, 133  
\bibitem{Bartinho} Cougo-Pinto M V, Farina C, Santos F C and Tort A C, {\it
The speed of light in confined QED vacuum: faster or slower than c?}.
Submitted to publication.
\bibitem{BORDAG} Bordag M, Robaschik D and
Wieczorek E, (1985) {\it Ann. Phys.} NY {\bf 165}, 192
\bibitem{HL} Heisenberg W and Euler H, (1936) {\it Z. f. Phys.} {\bf 98}
(1936), 714
\bibitem{Schwinger51} Schwinger J, (1951) {\it Phys. Rev.} {\bf 82}, 664
%
\bibitem{Grad} Gradshteijn I S and  Ryzhik I M, {\it Tables of Integrals,
Series and Products, Fifth Edition}, Academic Press, New York (1994)
%
\end{thebibliography}
\end{document}